\documentclass{article} 
\usepackage{spie}
\input{psfig}

\title{ALFA\,\&\,3D: integral field spectroscopy with adaptive optics}

\author{
R.I. Davies\supit{a},
M. Kasper\supit{b},
N. Thatte\supit{a},
M. Tecza\supit{a},
L.E. Tacconi-Garman\supit{a},\\
S. Anders\supit{a},
and
T. Herbst\supit{b}
\skiplinehalf 
\supit{a}Max-Planck-Institut f\"ur extraterrestrische Physik, 
Postfach 1603, 85740 Garching, Germany
\\
\supit{b}Max-Planck-Institut f\"ur Astronomie, 
K\"onigstuhl 17, 69117 Heidelberg, Germany
}

\authorinfo{Further author information: 
Send correspondence to R. Davies, email davies@mpe.mpg.de}

\def\spose#1{\hbox to 0pt{#1\hss}}
\def\simlt{\mathrel{\spose{\lower 3pt\hbox{$\mathchar"218$}}
     \raise 2.0pt\hbox{$\mathchar"13C$}}}
\def\simgt{\mathrel{\spose{\lower 3pt\hbox{$\mathchar"218$}}
     \raise 2.0pt\hbox{$\mathchar"13E$}}}
\def\arcsec{\nobreak{$''$}}
\newcommand{\mv}{\hbox{m$_{\rm V}$}}
\newcommand{\mk}{\hbox{m$_{\rm K}$}}
\newcommand{\av}{\hbox{A$_{\rm K}$}}
\newcommand{\um}{\,\hbox{$\mu$m}}
\newcommand{\kms}{\,\hbox{\hbox{km}\,\hbox{s}$^{-1}$}}
\newcommand{\msun}{\,\hbox{M$_{\odot}$}}

\begin{document}

\maketitle

%%%%%%%%%%%%%%%%%%%%%%%%%%%%%%%%%%%%%%%%%%%%%%%%%%%%%%%%%%%%%

\begin{abstract}
One of the most important techniques for astrophysics with adaptive
optics is the ability to do spectroscopy at diffraction limited scales.
The extreme difficulty of positioning a faint target accurately on
a very narrow slit can be avoided by using an integral field unit,
which provides the added benefit of full spatial coverage.
During 1998, working with ALFA and the 3D integral field spectrometer,
we demonstrated the validity of this technique by extracting and
distinguishing spectra from binary stars separated by only 0.26\arcsec.
The combination of ALFA\,\&\,3D is also ideally suited to imaging distant
galaxies or the nuclei of nearby ones, as its field of view can be
changed between 1.2\arcsec$\times$1.2\arcsec\ and
4\arcsec$\times$4\arcsec, depending on the pixel scale chosen.
In this contribution we present new results both on
galactic targets, namely young stellar objects, as well as
extra-galactic objects including a Seyfert and a starburst nucleus.
\end{abstract}


%%%%%%%%%%%%%%%%%%%%%%%%%%%%%%%%%%%%%%%%%%%%%%%%%%%%%%%%%%%%%

\section{INTRODUCTION}
\label{sec:intro}

The effort currently being directed at the development of adaptive
optics (AO) systems by the astronomical community is driven ultimately
by the science results they can bring.
To this end it is a requirement that spectroscopy can be done in
conjunction with AO.
Hitherto most results published have been pure broad-band imaging.
The reason is that classical spectroscopy imposes a number of
limitations, the most important of which concerns the slit width.
Either the slit is extremely narrow ($\simlt$0.1\arcsec) in 
which case it is difficult to align on faint targets, or it is wider
leading to loss of spatial resolution along that axis.
As a consequence mapping of sources on arcsec scales
becomes a very slow process.
Integral field spectroscopy overcomes both of these by obtaining
spectra of every pixel in a field that is large compared to the
diffraction limited PSF. 
Thus there is a reasonable degree of error permissible in positioning
faint targets, and the inner arcseconds of a source can be
simultaneously mapped in emission and absorption lines.

3D is an integral field spectrometer built by MPE (Weitzel et
al. 1996), which obtains H- or K- band spectra of an entire
16$\times$16 pixel field at resolutions of $R\sim1000$ or $R\sim2000$.
An Aperture Interchange Module (AIM, Anders et al. 1998) allows it to
be used with ALFA.
AIM provides 2 pixel scales: 0.07\arcsec\ giving a
1.2\arcsec$\times$1.2\arcsec\ field of view for the diffraction limited
mode, and 0.25\arcsec\ giving a `wide' 4\arcsec$\times$4\arcsec\ field
of view for partial correction (as might occur in poorer seeing
conditions).
Additionally AIM can switch between on-axis and off-axis fields so
that sky frames can be observed without opening the AO loop, a
necessary step in normal ALFA operation.
Serendipitously, although ALFA can only lock on a reference star that
is $\sim$20\arcsec\ off-axis, AIM allows stars much further
off-axis to be used.
In this mode, which was used for 2 of the objects presented here, the
telescope is pointed to a `sky' patch close to the reference star and
AIM's off-axis field is used to observe the target.

In this contribution we present a brief summary of some results we
have obtained using ALFA\,\&\,3D to observe both galactic
and extragalactic sources.
In Sections~\ref{sec:v1318cyg} and~\ref{sec:ttau} discuss the young stellar
objects V\,1318\,Cygni and T\,Tau.
We turn to the nucleus of the LINER galaxy NGC\,1161 in
Section~\ref{sec:ngc1161}, and finish in Section~\ref{sec:ngc1068} with
the archetypal Seyfert 2 nucleus of NGC\,1068.

%%%%%%%%%%%%%%%%%%%%%%%%%%%%%%%%%%%%%%%%%%%%%%%%%%%%%%%%%%%%%
\section{V\,1318\,Cygni}
\label{sec:v1318cyg}

\begin{figure}[ht]
\centerline{\psfig{file=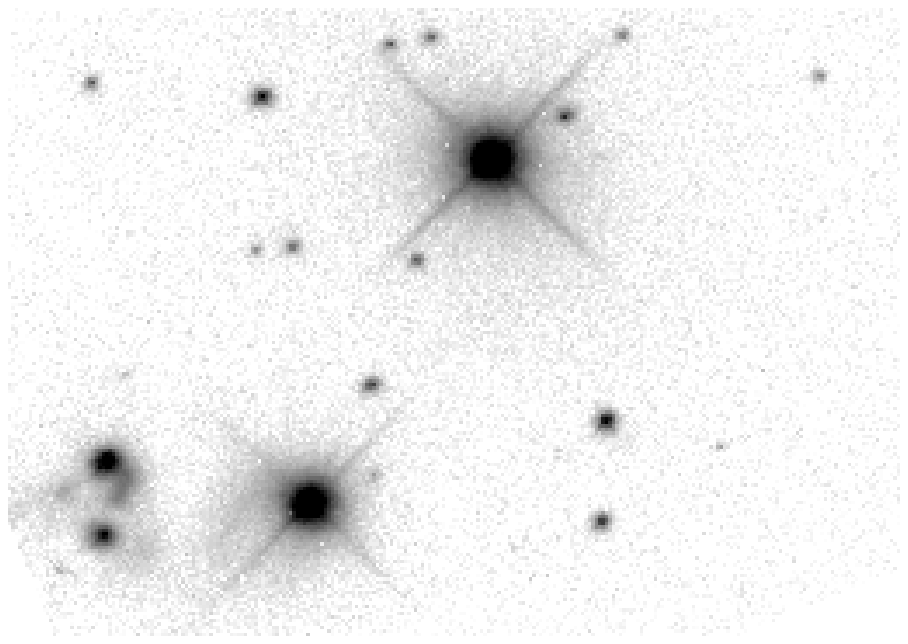,width=9cm}}
\vspace{-6.3cm}
\centerline{\psfig{file=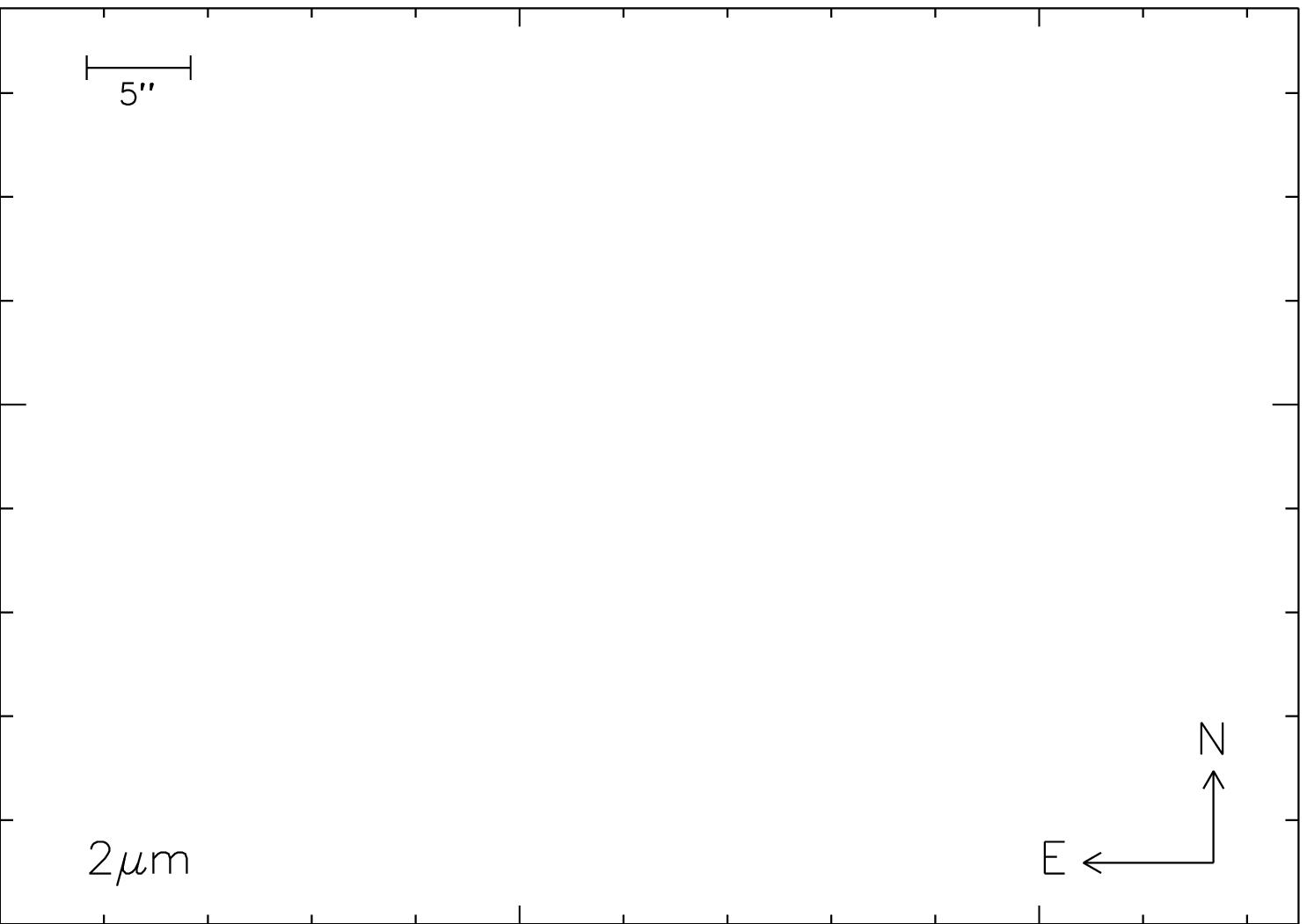,width=9cm}}
\caption{Field around BD\,+40$^\circ$\,4124 (top centre), plotted on a
logarithmic scale (from Davies et al. 1999).
The double system V\,1318\,Cygni is the pair to the lower left of the
image, with an obvious arm of continuum emission curling between them.
These stars are 36\arcsec\ from BD\,+40$^\circ$\,4124, which was used as
the wavefront reference.
\label{fig:bd40_field}}
\end{figure}

\subsection{Introduction}

The small stellar group around the two Herbig Ae/Be stars
BD\,+40$^\circ$\,4124 and V\,1686\,Cyg is a relatively isolated molecular
core containing a number of partially embedded low-mass stars
(Hillenbrand et al. 1995).
A K-band image of the field, observed using ALFA\,\&\,Omega-Cass in 1998
(Davies et al. 1999) with a {\sc fwhm} resolution of 0.25\arcsec, is
shown in Fig.~\ref{fig:bd40_field}.
In this Section we consider data obtained on
the pair of stars which constitute V\,1318\,Cyg (also known as
Lk\,H$\alpha$\,225), observed with
ALFA\,\&\,3D during Oct\,1999 using a pixel scale of 0.25\arcsec.
They are separated by 5.2\arcsec\ (so 2 overlapping fields were
required) equivalent to 5000\,AU at the distance of the
cluster (980\,pc, Hillenbrand et al. 1995).
While this is rather large, nearly 5\% of known YSOs do have
such wide separations (Mathieu 1994), although clearly they are not
likely to have a combined circumstellar envelope.
The wavefront reference, BD\,+40$^\circ$\,4124 (\mv=10.6,\mk=5.6),
lies 34\arcsec\ away and is the only star in the group brighter than
\mv=12.
The pixel scale, ambient seeing conditions, and distance to the
reference star limited the final resolution to $\sim$0.6\arcsec.

\begin{figure}[ht]
\vspace{5mm}

\centerline{
\psfig{file=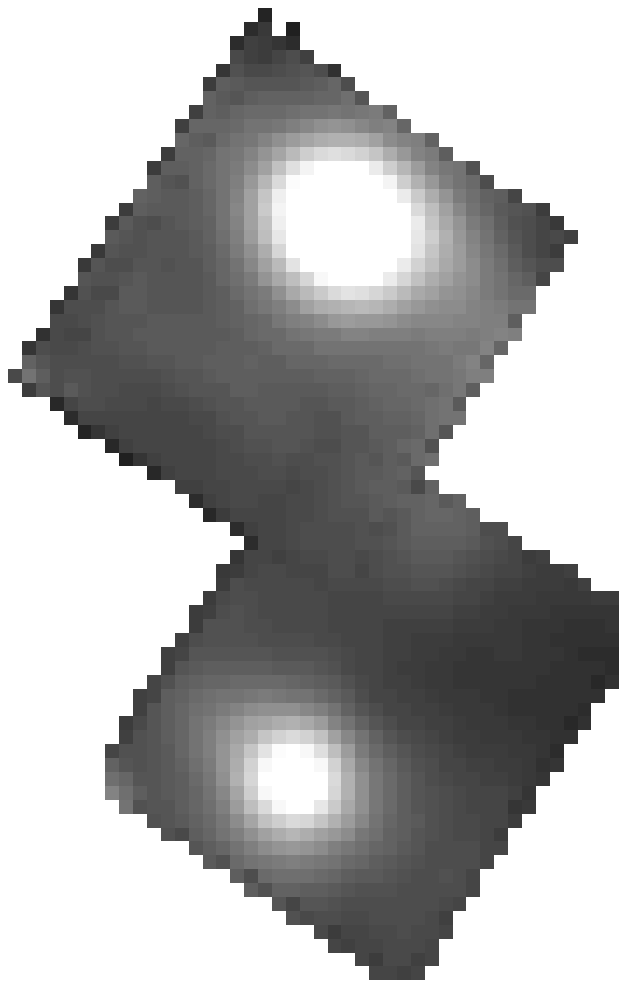,width=5cm}
\hspace{5mm}
\psfig{file=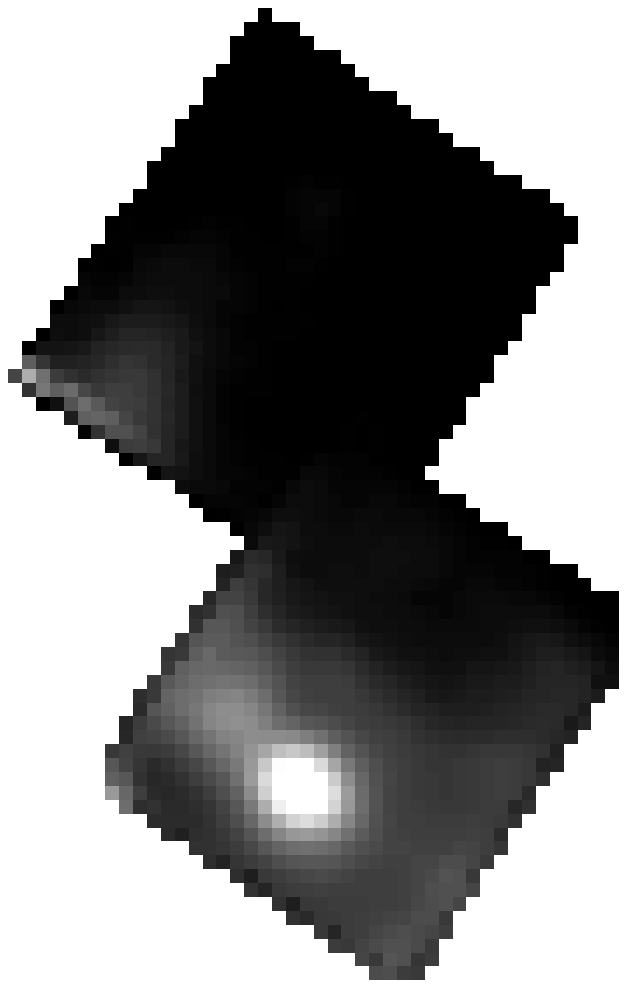,width=5cm}
\hspace{5mm}
\psfig{file=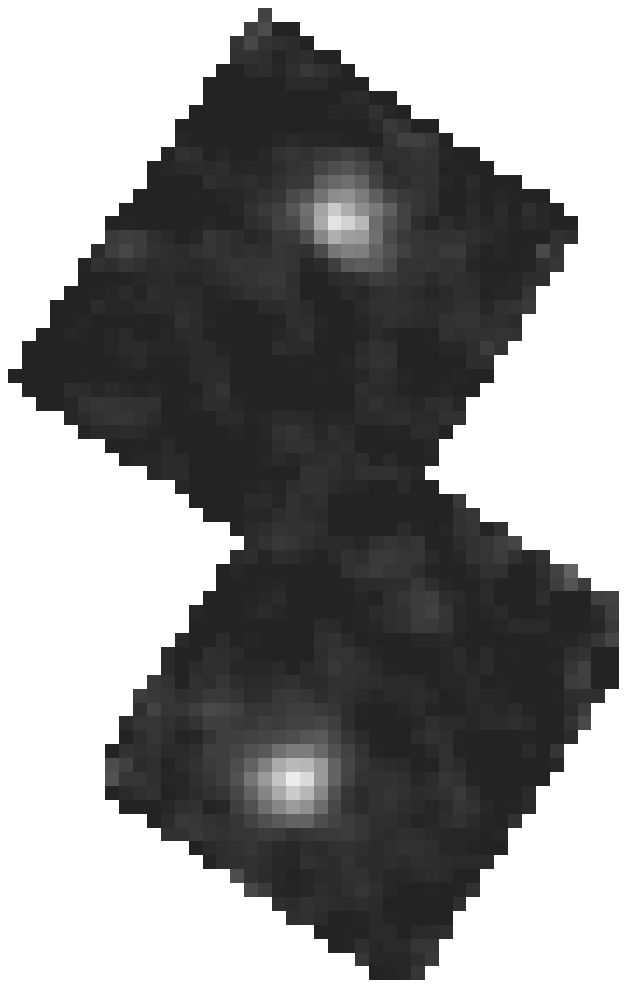,width=5cm}
}
\vspace{-7.8cm}
\hspace*{2.5mm}
\centerline{
\psfig{file=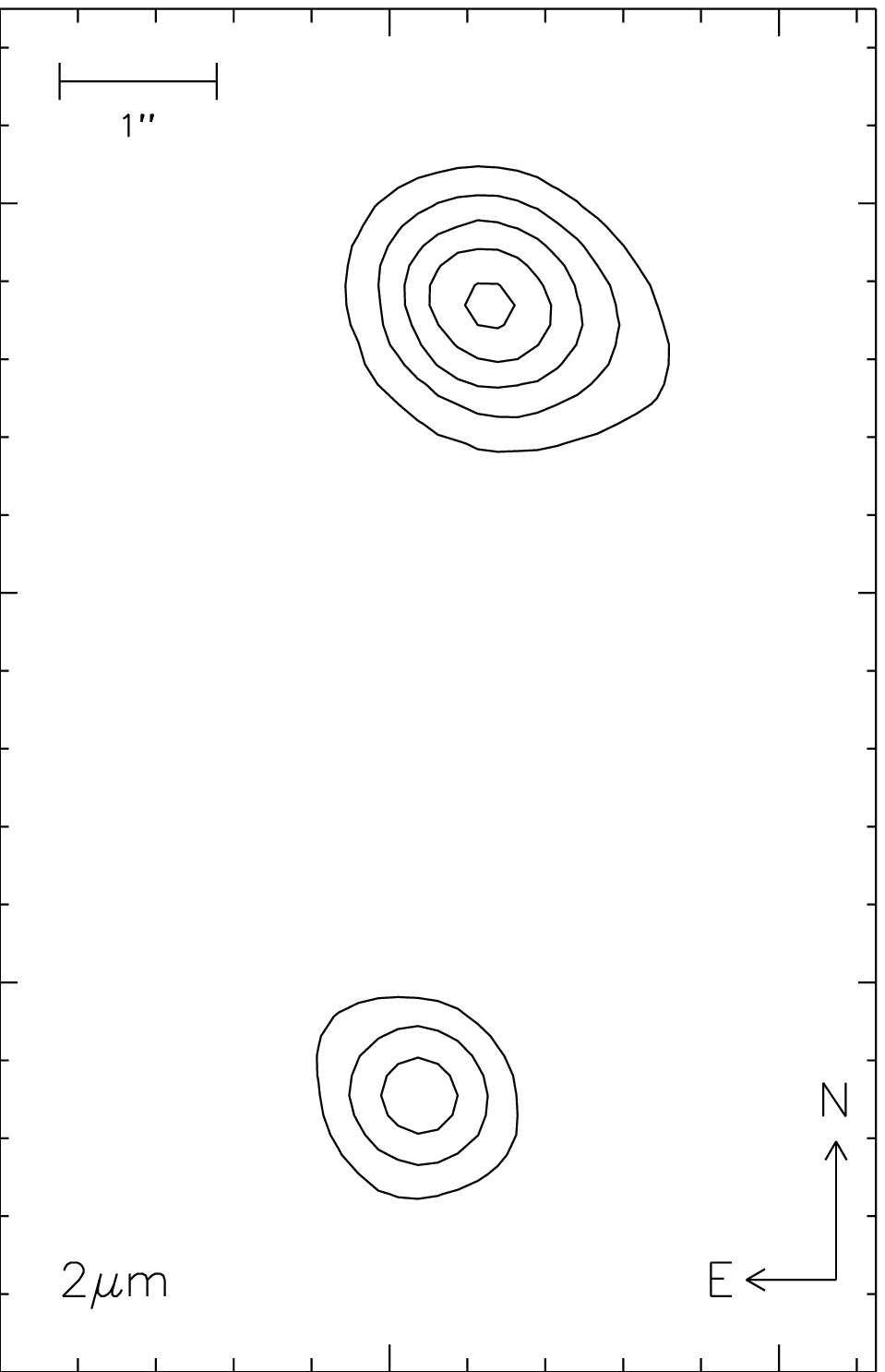,width=7.76cm}
\hspace{-22.6mm}
\psfig{file=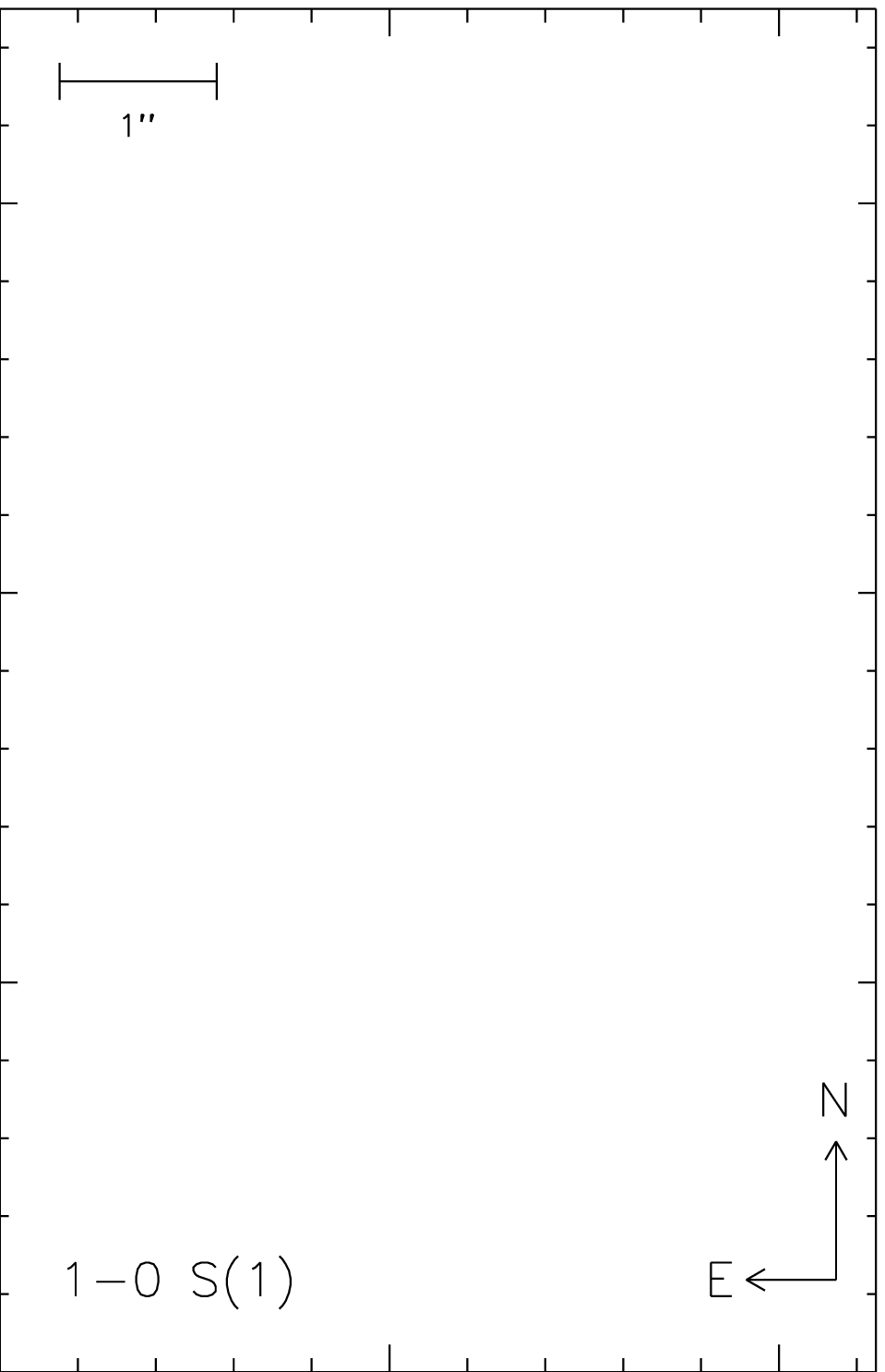,width=7.76cm}
\hspace{-22.6mm}
\psfig{file=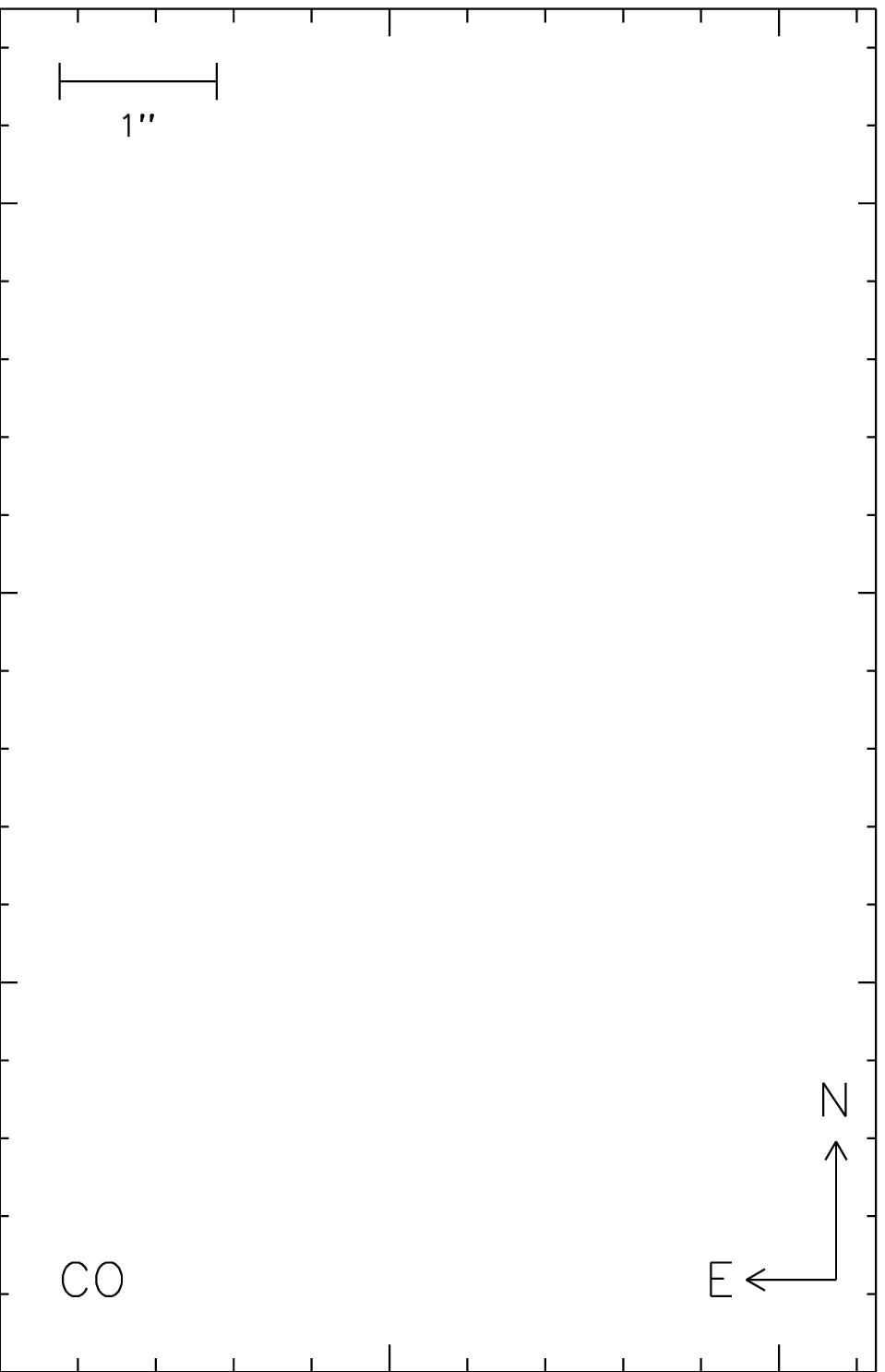,width=7.76cm}
\hspace{-22.6mm}
}

\caption{Images of V\,1318\,Cygni observed with 3D and ALFA.
Due to non-ideal observing conditions, and the
distance to the reference, the resolution achieved was $\sim$0.6\arcsec.
Left: K-band continuum showing (faintly) the emission curving to the
west between the stars.
Centre: H$_2$ 1-0\,S(1) line map -- there is strong emission both on
the southern star and to its east.
Right: CO emission map -- both stars have similar CO fluxas shown,
although the equivalent width for the northern star is much less.
\label{fig:bd40_imgs}}
\end{figure}

We present a continuum image as well as H$_2$ and CO emission line maps
in Fig~\ref{fig:bd40_imgs}, 
and plot spectra in Fig.~\ref{fig:bd40_spec}.
The spectra are rather different from the only other published K-band
spectra (Aspin, Sandell, \& Weintraub, 1994).
We find that the northern star is much brighter in K than the southern
star, and shows no evidence for H$_2$ emission.
On the other hand, both stars have a similar CO emission flux.
These characteristics are briefly discussed below.

\subsection{YSO Class}

\begin{figure}[ht]
\centerline{\psfig{file=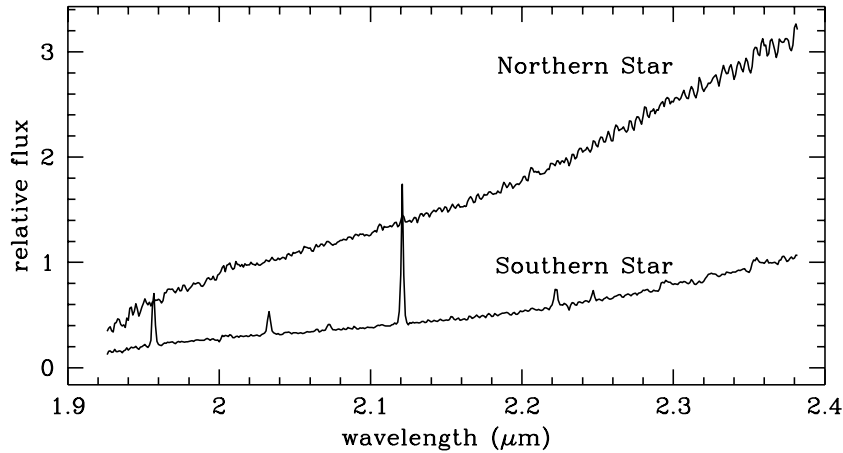,width=12cm}}
\caption{K-band spectra of the 2 stars in V\,1318\,Cygni, rather
different from the only other such spectra of V\,1318\,Cygni in the
literature (Aspin et al. 1994).
The southern star exhibits strong H$_2$ emission, while the northern
star has none.
The continuum in both cases is well matched by a 400--500\,K
blackbody and a 15--25\% contribution from an extincted A-type star.
\label{fig:bd40_spec}}
\end{figure}

The shape of the spectra are most similar to those of Class~I young
stellar objects presented in the Atlas of Greene \& Lada (1996).
These are the least evolved sources, surrounded by relatively massive
envelopes of gas and dust.
Class~I objects are associated with very high K-band veilings (the ratio of
excess to photospheric emission in the K-band, 
$r_{\rm K} = F_{\rm Kex}/F_{\rm K*}$). 
The excess emission is believed to come from hot dust or an accretion
disk around the star and does not necessarily redden the star itself.
Optical spectroscopy from Hillenbrand (1995) suggests that these stars
have spectra types mid~A to F, and extinctions of \av$>7$--8.
We have therefore fitted a simple model, comprising an extincted
(\av=10) mid to late A star (T$_{\rm eff}$=8000\,K) and a dust
component (with temperature T$_{\rm d}$ to be determined), to the
spectra.
The degeneracy that arises from varying the spectra type and extinction
does not permit a more general model to be used.
On the other hand fitting only a dust component leaves a significant
residual which rises towards the blue.
The northern star has best fitting parameters T$_{\rm d}$=520\,K and 
$r_{\rm K}$=3.4 ({\sc rms} residual 0.033, on the same scale as
figure~\ref{fig:bd40_spec});
the southern star yielded T$_{\rm d}$=430\,K and $r_{\rm K}$=1.7 
({\sc rms} residual 0.010).
This confirms the identification of V\,1318\,Cyg\,N as Class~I, but
suggests that V\,1318\,Cyg\,S is closer to Class~II in which the massive
envelope is dissipating or the accretion rate is slowing.

\subsection{Emission Line Features}

Fig~\ref{fig:bd40_imgs} (centre) is a map of the H$_2$ 1-0\,S(1)
emission, showing that the strongest emission is located directly on
the southern star and there is an outflow/jet or extra source to its east.
A Br$\gamma$ image is not shown as there is none detectable.
Although these observations are in stark contrast to those of T\,Tau
(Section~\ref{sec:ttau}), they do not contradict the survey of YSOs by
Greene \& Lada (1996) in which 23/44 of Class~I--II objects
showed Br$\gamma$ emission, and only 5/44 exhibited H$_2$ emission.

\begin{figure}[ht]
\centerline{\psfig{file=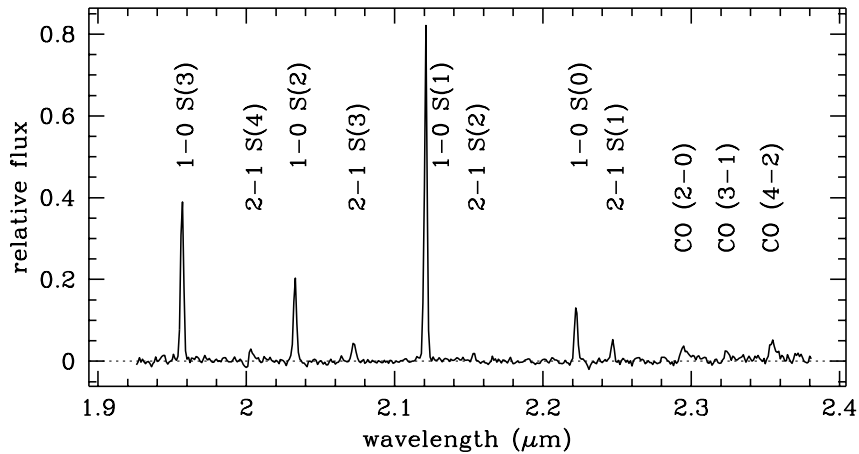,width=12cm}}
\caption{Continuum subtracted spectrum of the southern star of
V\,1318\,Cygni.
The same flux scaling as the previous figure is used.
8 H$_2$ emission lines in the $\nu$=1-0 and $\nu$=2-1 vibrational bands
are marked, as well as 3 CO bandheads.
There is no detectable Br$\gamma$ emission at 2.17\um.
\label{fig:bd40_lower}}
\end{figure}

A continuum-subtracted spectrum of the south star is presented in
Fig~\ref{fig:bd40_lower}, in which 8 H$_2$ lines are detectable at 
$\simgt$3$\sigma$, 4 from each of the $\nu$=1-0 and $\nu$=2-1 series.
The strength of the latter lines immediately rules out models of pure
thermal or fluorescent excitation.
Possible alternatives include a combination of these two, or excitation in
fast (100--300\kms) shocks leading to dissociation of the molecules.

CO emission has been known to occur in some YSOs for 20 years.
High spectral resolution observations suggest that emission from an
accretion disk is the most likely source, although it is not yet
possible to rule out neutral winds in all cases (Chandler, Carlstrom,
\& Scoville 1995).
In a survey of 40 YSOs, Carr (1989) observed 9 with CO emission.
For an accretion disk model he found that accretion rates of
10$^{-8}$-- a few $\times10^{-7}$\msun\,yr$^{-1}$ could account for
the CO bandhead fluxes and the correlation with stellar luminosity.
For a wind model mass loss rates of
10$^{-7}$--10$^{-5}$\msun\,yr$^{-1}$ were needed.
Greene \& Lada (1996) also considered accretion disks but in the
context of veiling, and found that accretion rates
$\sim5\times10^{-7}$\msun\,yr$^{-1}$ could account for 
$r_{\rm K}\simlt1$. 
But the greater accretion rates required for much higher veiling are
not compatible with CO emission.
It appears initially as if an accretion disk model could simultaneously
account for the veiling and CO emission in the southern source. 
In the northern source the CO flux is similar and could indicate
similar conditions (density, temperature, accretion rate) in the
accretion disk, but an additional hot dust envelope probably
contributes to the higher veiling.
This issue, and the others briefly touched upon above, will be examined
in more detail in a future paper.

%%%%%%%%%%%%%%%%%%%%%%%%%%%%%%%%%%%%%%%%%%%%%%%%%%%%%%%%%%%%%
\section{T\,Tauri}
\label{sec:ttau}

\subsection{Introduction}

T\,Tau is an unusually complex system of multiple (at least two)
stellar components, jets and outflows unlike that seen in any other T
Tauri star.  
The optically visible primary, T\,Tau\,N, was measured to
be a K1 star (Cohen \& Kuhi 1979), and has an companion 0.7\arcsec\ to the
south (Dyck, Simon \& Zuckerman 1982).  
This companion, T\,Tau\,S, exhibits a spectral energy distribution suggestive
of a very young embedded source (Koresko, Herbst \& Leinert 1997)
assigning it to the class of infrared companions (IRC) of
which very few are known today.  
It shows an apparent lack of cold dust emission and has undergone a
brightness flare of 2 magnitudes at various wavelengths between 1989
and 1991 (Ghez et al. 1991).  
Recent speckle holography observations revealed
an companion to T\,Tau\,S which is separated by only 0.05\arcsec\ 
(Koresko, 2000).  
The near environment of T\,Tau is a source of
surprisingly strong extended 2.12\um\ $H_2$ $\nu$=1-0 S(1)
ro-vibrational line emission (Beckwith et al. 1978). 
The spatial distribution of this excited $H_2$ shows two separate outflows
which are perpendicular to each other in projection 
(Langevelde et al. 1994, Herbst et al. 1996).
Recent optical spectroscopy (Solf \& B\"ohm 1999)
indicated that these outflows are in fact
almost perpendicular with the N-S component originating
from T\,Tau\,S and the E-W component originating from T\,Tau\,N. This
supports the model that T\,Tau\,S is intrinsically similar to T\,Tau\,N
but is obscured by the outer parts of T\,Tau\,N's disk 
(Hogerheijde et al. 1997).

Investigating these small angular scale phenomena is a formidable
observational challenge, since the seeing limit from ground-based
observatories smears detail on scales of approximately hundred AU.
High angular resolution for imaging and spectroscopy is mandatory for
untangling these phenomena.  
Diagnostic spectral lines include the Br$\gamma$ (n=7-4) 2.17\um\ 
recombination line of hydrogen, which traces
stellar winds, ionized regions and accretion, and the 2\um\ 
quadrupole lines of molecular hydrogen, which, through their
distribution and ratios, will point to regions of shock excitation and
ultraviolet fluorescence.

\begin{figure}[ht]
\centerline{\psfig{file=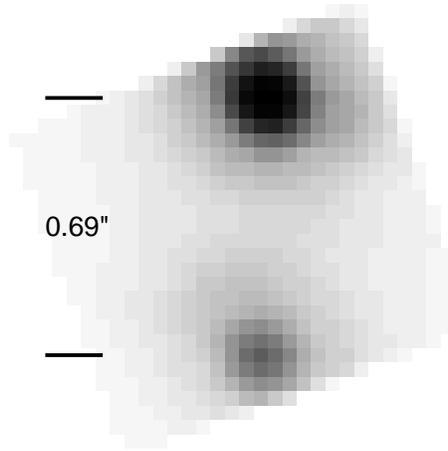,width=60mm}}
\caption{K-band continuum image of the T\,Tauri binary,
displaying the whole field of view of 3D of approximately 
1 square arcsecond.
The image has been rotated so that North is up and East is left.
\label{fig-ttauIma}}
\end{figure}

\begin{figure}[ht]
\centerline{\psfig{file=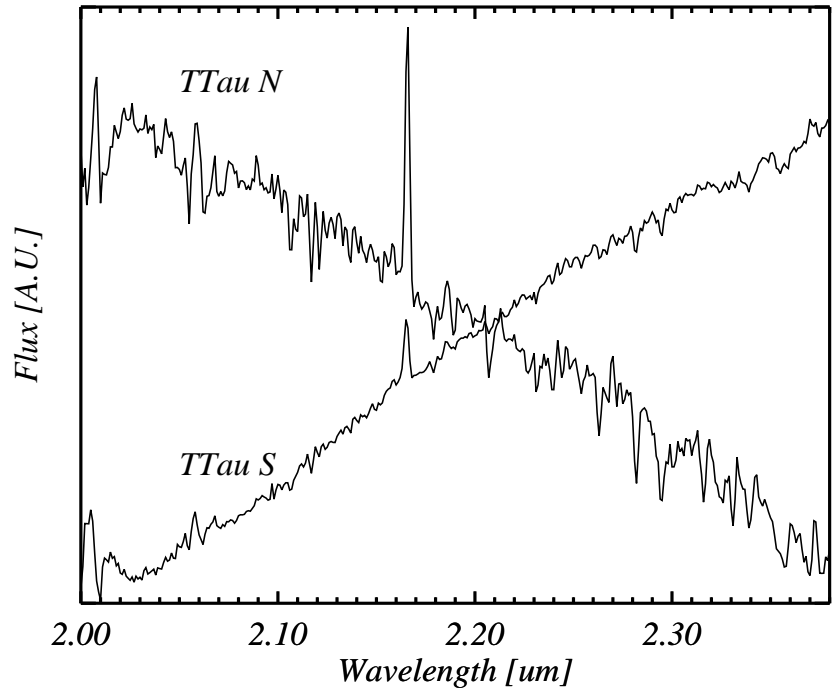,angle=90,width=130mm}}
\caption{K-band spectrum of T\,Tau\,N and T\,Tau\,S with the left scale 
corresponding to the spectrum of the primary, and the right scale to
the secondary.
The apparent helium emission  at 2.05\um\ is in fact due to imperfect
removal of telluric features.
The only obvious emission line is Br$\gamma$, and neither star shows
any H$_2$ emission.
\label{fig-ttKspec}}
\end{figure}

\subsection{Observations} 

T\,Tauri was observed with the 3D imaging spectrometer in H- and K-band
in October 1999 at the 3.5-m-telescope on Calar Alto with a spectral
resolution of $R \approx 2100$ in H and $R \approx 1100$ in K. 
The adaptive optics system ALFA used T\,Tau\,N (\mv$=9.6$) as a natural
guide star delivering angular resolutions of 0.14\arcsec\ in H and
0.16\arcsec\ in K. 
The total integration time was 14~and 18~minutes respectively .

Standard techniques of sky substraction and flat fielding removed 
background emission, dark current and pixel-to-pixel gain
variations. 
A G2V star was observed at similar airmass to T\,Tau
and served as a photometric, spectroscopic and PSF reference. 
Division by the spectroscopic standard and subsequent
multiplication by a solar spectrum of the same spectral resolution
eliminated telluric features,  photospheric features of the G2V star,
and produced the correct spectral slope across the band. 
The final data structure consists of 600 separate wavelength slices, each
containing the spatial information  over a 1\arcsec$\times$1\arcsec\ 
region.

Fig.~\ref{fig-ttauIma} shows a K-band continuum image of the inner
square arcsecond of T\,Tau which is the whole high resolution
field of view of 3D when used with ALFA. T\,Tau\,S, which was reported
to be extended compared to T\,Tau\,N (Roddier et al. 1998), is
point-like in the
K-band continuum and in the Br$\gamma$ line-map.  The H-band data,
which is of a somewhat better resolution, indicates that T\,Tau\,S is
slightly elongated at a position angle corresponding to the
rescently discovered companion. 
Both stars are still unresolved in the Br\,10 (1.74\um) line-map.

Fig.~\ref{fig-ttKspec} displays the K-band spectra of T\,Tau\,N and T\,Tau\,S. 
The most prominent feature is the Br$\gamma$ emission line with
equivalent widths of 2.4\,\AA\ for T\,Tau\,N and 1.5\,\AA\ for T\,Tau\,S. 
While earlier observations showed Br$\gamma$ to be
confined to T\,Tau\,N (Herbst et al. 1996), our data indicates that both
components are accreting classical T\,Tauri stars.  
Furthermore the stars themselves do not appear to be responsible for
much H$_2$ emission,  a large amount of which was found
in the vicinity of T\,Tau at seeing scales (Herbst et al. 1996).
The photospheric features of T\,Tau\,N are consistent with a
veiled ($r_{\rm k} \approx 1$) spectrum of a K1 star, while T\,Tau\,S
appears almost featureless.\\

A detailed analysis of all the H- and K-band data will be the subject
of a future paper (Kasper, Herbst, Looney in prep).

%%%%%%%%%%%%%%%%%%%%%%%%%%%%%%%%%%%%%%%%%%%%%%%%%%%%%%%%%%%%%
\section{NGC\,1161}
\label{sec:ngc1161}

\subsection{Introduction}

Optical spectroscopy of NGC\,1161 by Ho, Filippenko, \& Sargent (1997)
suggested that NGC\,1161 is a `transition object', exhibiting
properties similar to both LINER and starburst. In particular the
high ratios [N{\sc ii}]/H$\alpha$=1.75, [S{\sc ii}]/H$\alpha$=0.52, and
[O{\sc i}]/H$\alpha$=0.14 tend to put the galaxy in the AGN region of
the diagnostic diagrams developed by Fruscione \& Griffiths (1991).
Furthermore Ho et al. claim to have detected a broad component to the
H$\alpha$ emission, constituting about half the total H$\alpha$ flux.

\begin{figure}[ht]
\vspace{3mm}

\centerline{
\psfig{file=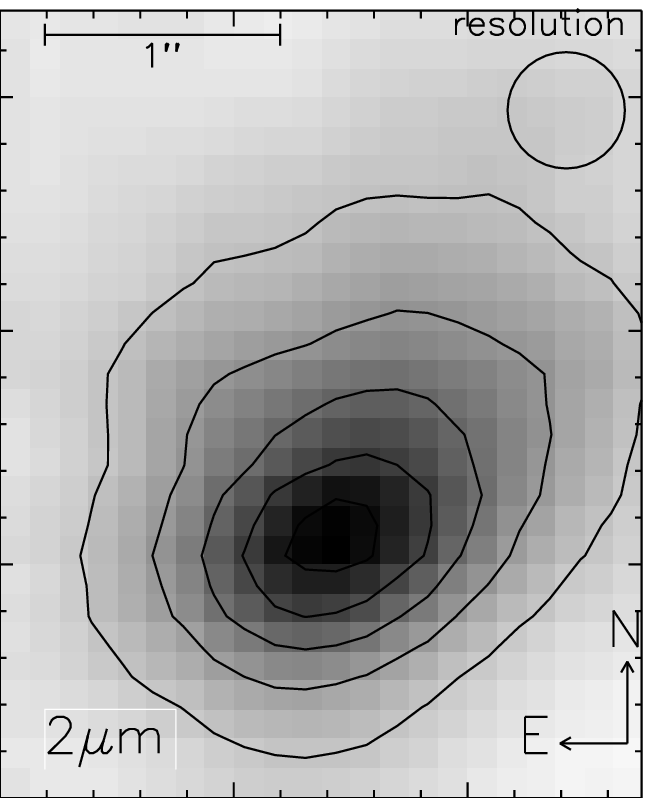,width=5cm}
\hspace{5mm}
\psfig{file=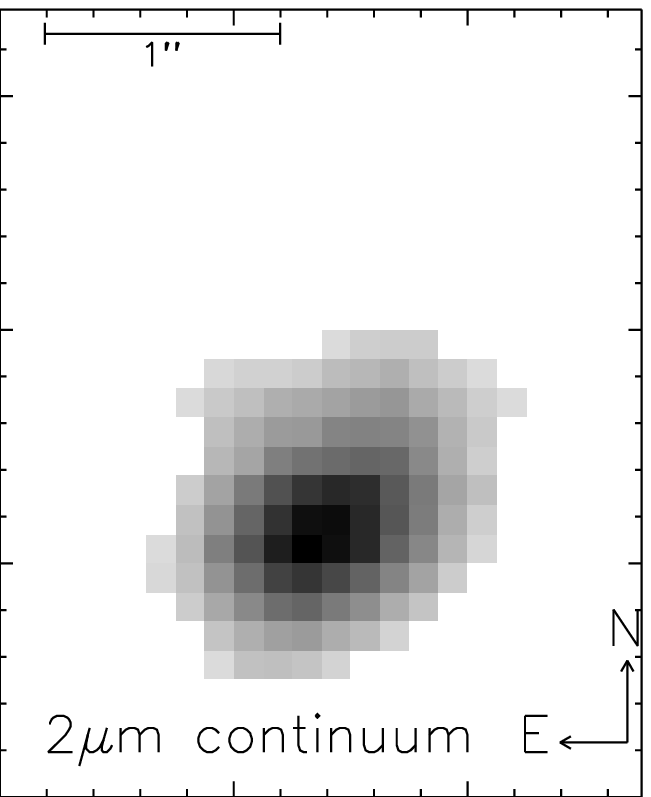,width=5cm}
\hspace{5mm}
\psfig{file=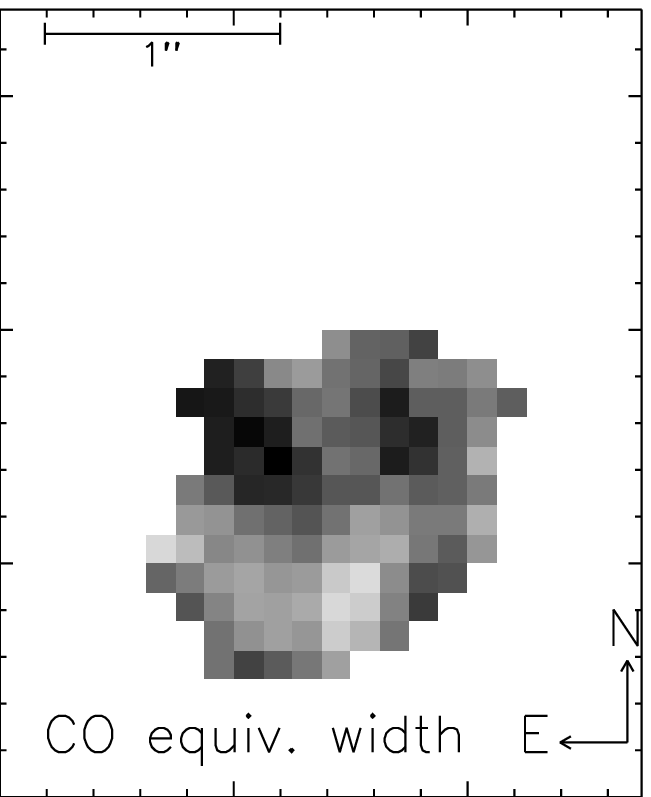,width=5cm}
}
\caption{Images of NGC 1161 at a pixel scale of 0.125\arcsec\ and a
resolution of about 0.5\arcsec\ (the wavefront reference star was
45\arcsec\ off-axis).
Left: full (approx. 3\arcsec$\times$3\arcsec) K-band continuum image.
Centre: masked image which includes only pixels with high signal to
noise in the spectrum.
Right: CO equivalent width map -- the deepest CO absorption occurs in 2
distinct regions away from the nucleus.
\label{fig:ngc1161_img}}
\end{figure}

In order to shed light on the excitation source in the nucleus of this
galaxy, we observed it with ALFA\,\&\,3D in Oct~99.
The wavefront reference was the \mv=9.9 star PPM\,45763 45\arcsec\ to
the east.
A pair of stars from the Washington Double Star Catalogue was found
which matched this configuration, allowing the PSF to be measured.
As a result of the large offset and also due to the non-ideal seeing,
the resolution achieved was 0.5\arcsec.
Fig~\ref{fig:ngc1161_img} (left) shows the K-band image of the galaxy,
in which the nucleus is clearly extended on sub-arcsec scales.

\subsection{Stellar Populations}

The central image in Fig.~\ref{fig:ngc1161_img} has been masked,
leaving only those pixels where there is good signal to noise in the
K-band continuum.
In these pixels the equivalent width of the CO absorption in the first
2 bandheads has been measured and is shown in the right-hand image.
This shows that the deepest CO absorption -- corresponding to the
largest fraction of late-type supergiant stars or least dilution form
hot dust -- occurs in 2 regions distinct from the continuum peak.

\begin{figure}[ht]
\centerline{\psfig{file=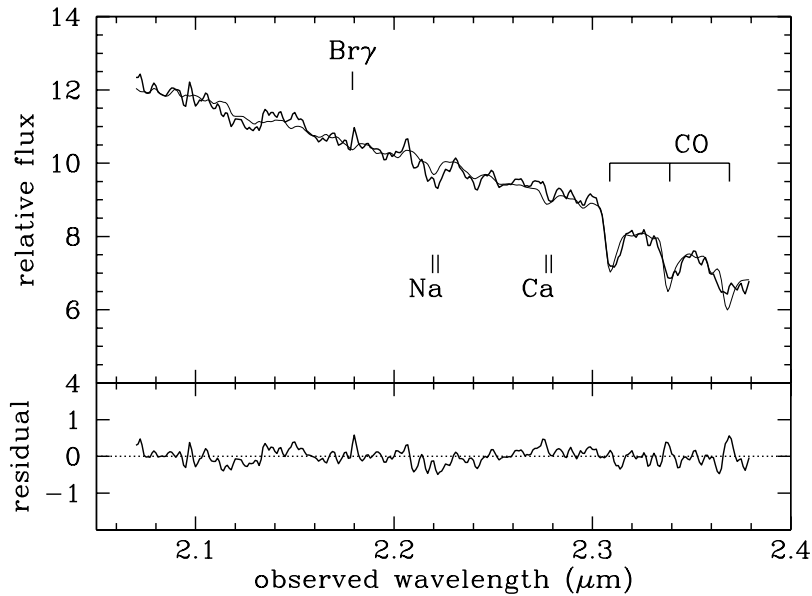,width=11cm,angle=-90}}
\caption{K-band spectrum of the inner 1\arcsec\ of NGC 1161, at the
observed wavelength.
The spectrum has been fitted with late-type templates (both giant and
supergiant) and the difference spectrum is shown underneath.
\label{fig:ngc1161_spec}}
\end{figure}

A spectrum of the inner 1\arcsec\ is presented in
Fig~\ref{fig:ngc1161_spec} (spectra of smaller individual regions are
similar).
Overdrawn is a best-fit spectrum made from templates of late-type
stars, consisting of 10\% M-type supergiants and 90\% K0 and K5 giants.
Fitting only the wavelength range 2.25--2.37\um\ gives a 25\%
contribution from supergiants.
The existence of supergiants indicates that star formation has occured
in at least 2 sites in the last 10$^7$\,yr.
It is interesting that while the first CO bandhead is very well fit,
the second is less so and the third not at all.
This may suggest a contribution of cool $\sim$500\,K dust to the
emission at longer wavelengths.

The spectrum shows that any activity in the continuum peak is dominated
by a stellar population typical of normal galaxies.
Since the flux of late-type stars peaks in the infrared, 
it is natural to look at the optical spectrum to see if there is a
population of bluer OB stars. 
The spectrum of Ho et al. extracted in a 2\arcsec$\times$4\arcsec\ box
also shows little evidence for young stars.
A fit including A stars is a marginally better fit than one without,
but a UV spectrum would be needed to clarify the issue.

Our conclusion is that the main episodes of star formation on NGC\,1161
are over, so that the only remnants observable in the infrared are
late-type supergiants.
In the optical, the narrow component of H$\alpha$ may be produced by A
stars which have not yet evolved away from the main sequence.
The broad H$\alpha$ may suggest that there is a galactic wind resulting
from supernovae, and other lines such as [N{\sc ii}] and [S{\sc ii}]
may be excited in shocked clouds as this wind expands.
If this is the case H-band observations of [Fe{\sc ii}] at 1.64\um,
which traces supernova remnants in starbusts, would help to explain the
observations.

%%%%%%%%%%%%%%%%%%%%%%%%%%%%%%%%%%%%%%%%%%%%%%%%%%%%%%%%%%%%%
\section{NGC 1068}
\label{sec:ngc1068}

\subsection{Introduction}

NGC\,1068 holds a special interest both in astrophysics, because it is so
close and is considered an archetypal Seyfert 2 galaxy, and also in
adaptive optics, because it is faint and in the optical is extended.
Correcting on the nucleus of NGC\,1068 has been likened to earning the
`driving licence of AO systems'.

For a Shack-Hartmann sensor we can consider the number of photons
available in each spot on which it is necessary to estimate the
centroid.
To do so we assume 70\% atmospheric transmission, 70\% telescope and
instrument transmission (including detector quantum efficiency), 50\,cm
subapertures (matched to 1\arcsec\ seeing), 100\,Hz frame rate.
For NGC\,1068, in a 2.8\arcsec\ aperture \mv=13.5 (Antonucci \& Miller,
1985) while \mk=10.1 (Thatte et al. 1997).
Hence there are only $\sim$200 photons for the centroid calculation, rather
too few for a robust estimation (Kasper et al. 1999).
To obtain a good result requires either excellent seeing, but partial
correciton can be achieved by undersampling the wavefront.

ALFA was able to lock on this galaxy for the first time in Oct~99 and
achieve a resolution of 0.25\arcsec\ in the H-band, confirming that in
this waveband the nucleus is single and has no other bright knots of
emission nearby.
More recently, it has been possible to correct on fainter objects and
during Feb~00 it was possible to use Mkn\,231 as the wavefront
reference (integrated \mv=13.6, and \mv$\sim$14.3 in the nucleus).

\begin{figure}[ht]
\centerline{
\psfig{file=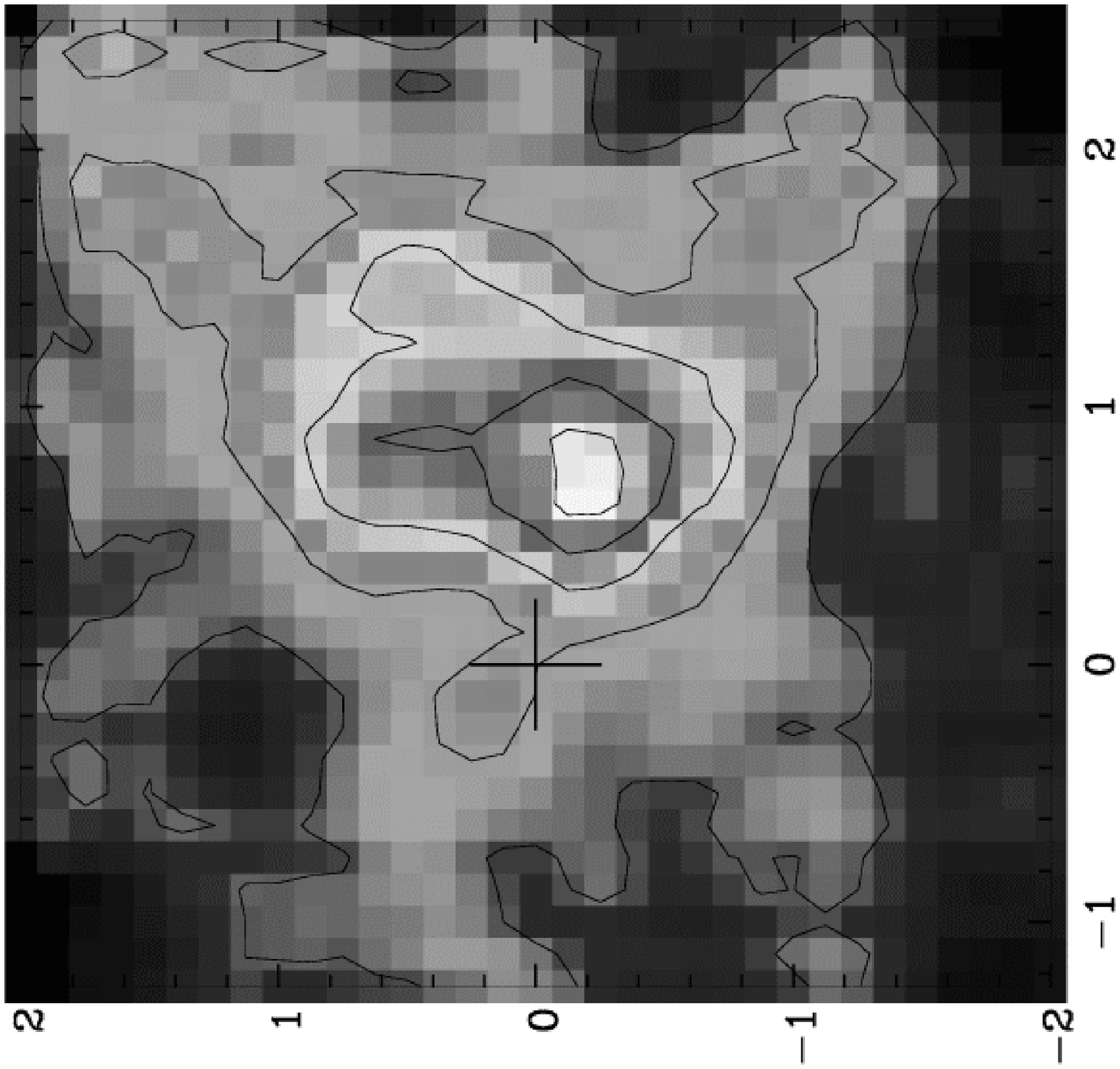,width=8cm,angle=30}
%\hspace{5mm}
%\psfig{file=ngc1068_co.eps,width=8cm,angle=30}
}
\vspace{-1cm}
\centerline{\hspace*{3cm}[Fe {\sc\,ii}] 1.64\um
%\hspace{7cm} CO 1.62\um
}
%\vspace{0.5cm}
\caption{[Fe\,{\sc ii}] 1.644\um line map of NGC 1068:
The image has been rotated so that North is up and East left, and
angular offsets from the continuum nucleus (marked by a cross) are
given in arcsec.
In the smallest pixel scale, a resolution of $\sim$0.25\arcsec\ was
reached.
But to cover the extended emission out to 150\,pc, the coarser pixel
sampling was used for the observations presented here, 
limiting the resolution to $\sim$0.5\arcsec.
\label{fig:ngc1068}}
\end{figure}

\subsection{Line Map}

The first 2\um\ integral field spectroscopy of NGC\,1068 was presented
by Blietz et al. (1994).
At a resolution of 1\arcsec, they found that the 1.64\um\ [Fe{\sc ii}]
emission peaked close to the continuum nucleus and was extended
northeast along the direction of the radio jet.
The surface brightness of the line emission was greatly reduced where
the radio continuum widened.
Their interpretation was that the line emission arises at the interface
bewteen the nuclear outflow and surrounding molecular clouds.
The [Fe{\sc ii}] could be excited either by X-ray photo-ionisation of
the clouds, or as a result of fast J-shocks destroying grains which
contain most of the interstellar iron.

Fig~\ref{fig:ngc1068} shows the new image from ALFA\,\&\,3D
which reveals more structure in the [Fe{\sc ii}] line.
The peak of the line emission is to the north of the continuum nucleus,
and there is some faint emission to the south.
Examination of the spectra suggest that there is a difference in
line-of-sight velocity between the north and south emission.
A detailed analysis of the [Fe{\sc ii}] line and other data will be
presented elsewhere (Thatte et al. in prep).

\section{Conclusion}

We have presented a selection of results obtained using the ALFA
adaptive optics system and the 3D integral field spectrometer during
October~1999.

Although only one of the results is at the diffraction limit of the
telescope the others demonstrate that AO is still viable in the
partial correction mode, to improve poor seeing or when using far
off-axis reference stars.

Together they show the real advantages of integral field techniques over
conventional longslit spectroscopy when used in conjunction with
adaptive optics.
However, one disadvantage is also abundantly clear:
the spatial field is very limited if one dimension of the detector is
used for the spectral regime.
Here we used a 256$^2$ detector for a 3.5-m telescope (at Calar Alto), but
the problem will remain for 8-m telescopes even with larger detectors
due to the smaller diffraction limit.
Nevertheless, these preliminary results are very encouraging for the
future of integral field spectroscopy with adaptive optics.

%%%%%%%%%%%%%%%%%%%%%%%%%%%%%%%%%%%%%%%%%%%%%%%%%%%%%%%%%%%%%

\acknowledgments

The authors extend their thanks to everyone who has helped in this
project with their hard work and enthusiasm. 
In particular we gratefully acknowledge the commitment of the ALFA
team, the 3D team, and the staff at Calar Alto.

%%%%%%%%%%%%%%%%%%%%%%%%%%%%%%%%%%%%%%%%%%%%%%%%%%%%%%%%%%%%%
%\bibliography{report}
\bibliographystyle{spiebib}

%%%%%%%%%%%%%%%%%%%%%%%%%%%%%%%%%%%%%%%%%%%%%%%%%%%%%%%%%%%%%

\end{document}